\documentclass[aip,jcp,amsmath,amssymb,reprint]{revtex4-1}

\usepackage[english]{babel}
\usepackage{graphicx}
\usepackage{dcolumn}
\usepackage{booktabs}
\usepackage[load=abbr,alsoload=synchem]{siunitx}
\usepackage[colorlinks,allcolors=black,citecolor=blue,urlcolor=blue]{hyperref}

\graphicspath{{figures/}}

\begin{document}

\title{Unravelling the influence of quantum proton delocalization on electronic charge transfer through the hydrogen bond}

\author{Christoph Schran}
\affiliation{Lehrstuhl f\"ur Theoretische Chemie, Ruhr-Universit\"at Bochum, 44801 Bochum, Germany}
\affiliation{Department of Chemistry, Stanford University, Stanford, California 94305, USA}

\author{Ondrej Marsalek}
\affiliation{Department of Chemistry, Stanford University, Stanford, California 94305, USA}

\author{Thomas E. Markland}
\email{tmarkland@stanford.edu}
\affiliation{Department of Chemistry, Stanford University, Stanford, California 94305, USA}

\date{\today}

\begin{abstract}
Upon hydrogen bond formation, electronic charge density is transferred between the donor and acceptor, impacting processes ranging from hydration to spectroscopy.
Here we use ab initio path integral simulations to elucidate the role of nuclear quantum effects in determining the charge transfer in a range of hydrogen bonded species in the gas and liquid phase. 
We show that the quantization of the nuclei gives rise to large changes in the magnitude of the charge transfer as well as its temperature dependence. We then explain how a single geometric parameter determines the charge transfer through the hydrogen bond.
These results thus demonstrate that nuclear quantum effects are vital for the accurate description of charge transfer and offer a physically transparent way to understand how hydrogen bonding gives rise to it.
\end{abstract}


\maketitle

\section{Introduction}

The transfer of electronic charge density through a hydrogen bond manifests in effects ranging from the stabilization and spectroscopy of hydrated ions and aqueous solutions~\cite{Zhao2010/10.1063/1.3283900,Ben-Amotz2011/10.1021/jz2002875,Hynes2000/10.1021/ja993058q,Torii2014/10.1021/ct4011147} to the charging of polar residues in proteins~\cite{Martinez2011/10.1021/jz200697c}, and has even been implicated in explaining the observed zeta potential at hydrophobic interfaces~\cite{Takahashi2005/10.1021/jp0445270,Vacha2011/10.1021/jz2014852,Ben-Amotz2011/10.1021/jz2002875}. However, defining the amount of charge transfer (CT) remains a subject of significant debate, since partitioning the electron density and assigning it to particular atoms or molecules is not uniquely defined. This has led to the introduction of a number of approaches to calculate the charge transfer and charge transfer energy, with different approaches exhibiting both different quantitative and qualitative behavior~\cite{Mulliken1955/10.1063/1.1740588, Hirshfeld1977/10.1007/BF00549096, Reed1985/10.1063/1.449486, Stevens1987/10.1016/0009-2614(87)80143-4, Breneman1990/10.1002/jcc.540110311, Bader1991/10.1021/cr00005a013, Stone1993/10.1016/0009-2614(93)80058-W, Lillestolen2009/10.1063/1.3243863, Azar2012/10.1063/1.3674992}. For example, methods based on natural bond orbitals suggest that without CT the 0~K water dimer would adopt a different structure~\cite{Reed1983/10.1063/1.445134,Glendening2005/10.1021/jp058209s}. However, studies using other methods suggest a much less substantial role of CT in stabilizing the water dimer hydrogen bond~\cite{Khaliullin2009/10.1002/chem.200802107,Ronca2014/10.1002/cphc.201402321,Stone2009/10.1016/j.cplett.2009.03.073, Stone2017/10.1021/acs.jpca.6b12930}.

Previous CT studies of hydrogen bonded systems have analyzed minimum energy structures~\cite{Glendening2005/10.1021/jp058209s, Khaliullin2009/10.1002/chem.200802107,Ronca2014/10.1002/cphc.201402321} or those obtained from simulations that treat the nuclei classically~\cite{Lee2011/10.1063/1.3589419, Vacha2011/10.1021/jz2014852}.
However, nuclear quantum effects (NQEs), such as zero-point energy and tunnelling, are known to significantly change the geometry of hydrogen bonded systems~\cite{Marx1997/10.1126/science.275.5301.817,Marx1999/10.1038/17579,Car2008/10.1103/PhysRevLett.101.017801,Markland2012/10.1088/0953-8984/24/28/284126,Ceriotti2013a/10.1073/pnas.1308560110,Wang2014/10.1063/1.4894287}.
The zero-point energy along the hydrogen bond increases the ability of the proton to delocalize between the donor and acceptor, resulting in shorter and stronger hydrogen bonds.
On the other hand, the distortion of the hydrogen bond due to quantization of the librational modes perpendicular to the hydrogen bond leads to weaker hydrogen bonding.
The relative importance of these two competing quantum effects~\cite{Manolopoulos2009/10.1063/1.3167790} leads to cases where NQEs strengthen~\cite{Michaelides2011/10.1073/pnas.1016653108} or weaken~\cite{McKenzie2014/10.1063/1.4873352} hydrogen bonds, depending on the geometry and chemical environment. This competition has been used to elucidate the seemingly anomalously small NQEs and resulting isotope effects observed in many systems~\cite{Ceriotti2016/10.1021/acs.chemrev.5b00674}.

Here we investigate the role of NQEs in modulating the CT characteristics of a range of hydrogen bonded systems involving water and ions in both the gas and condensed phase and show how these can be understood based on the changes in the hydrogen bond geometry. We demonstrate that NQEs give rise to large changes in both the amount of CT and its contribution to the donor-acceptor interaction energy, independent of the method used to study charge transfer. In particular, including NQEs increases CT in some cases, while decreasing it in others. It also leads to a much milder temperature dependence of CT in all cases. To uncover the reasons for this, we show that the amount of charge transfer is almost exclusively determined by a scaled proton sharing coordinate. We use this observation to demonstrate that the charge of a water molecule in liquid water can be accurately estimated from the values of this coordinate for each hydrogen bond the molecule forms, thus provides a straightforward approach to modelling CT in solution.

\section{Simulation Details}

To evaluate NQEs on electronic properties in the gas phase, \SI{20}{\pico\second} ab initio path integral molecular dynamics (AI-PIMD) simulations at \SI{25}{\kelvin} and \SI{300}{\kelvin} of the water dimer as well as the water-fluoride and water-iodide complex were performed. These simulations were performed with a time step of \SI{0.5}{\femto\second} with 256 and 32 replicas, at the two respective temperatures.
The B3LYP~\cite{Becke1988/10.1103/PhysRevA.38.3098,Lee1988/10.1103/PhysRevB.37.785} exchange correlation functional and the \mbox{6-311++G**} Pople basis set~\cite{Pople1980/10.1063/1.438955,Clark1983/10.1002/jcc.540040303} for the water dimer and fluoride-water complex as well as the def2-SVPD basis set~\cite{Peterson2003/10.1063/1.1622924,Rappoport2010/10.1063/1.3484283} for the iodide-water complex were used to calculate electronic interactions on the fly during the simulations. An additional \SI{1}{\pico\second} at the beginning of each simulation was discarded as equilibration after starting from the optimized minimum energy configuration. Langevin thermostats~\cite{Leimkuhler2013/10.1093/amrx/abs010} with optimal coupling to each staging mode~\cite{Ceriotti2010/10.1063/1.3489925} were used to sample the canonical distribution. Two independent \SI{21}{\pico\second} classical ab initio molecular dynamics (AIMD) simulations of each system were carried out using the same electronic structure method, time step and temperature, together with massive Nos\'{e} Hoover chains thermostatting~\cite{Martyna1996/10.1080/00268979600100761} with a chain length of 6. The simulations were performed with the in-house Python program PyMD~\cite{Marsalek2014/10.1021/ct400911m} that uses IPython parallelization~\cite{Granger2007/10.1109/MCSE.2007.53} and the Atomic Simulation Environment~\cite{Jacobsen2002/10.1109/5992.998641} to link to Gaussian 09~\cite{g09} for the electronic structure calculation.
Our choice of basis set and number of replicas gives good convergence for geometrical properties (see SI).
Since it is known that density functional theory (DFT) can cause spurious charge delocalization effects~\cite{Yang2008/10.1103/PhysRevLett.100.146401}, we validated our choice of methods by also conducting MP2 simulations of the water dimer, as detailed in the SI.
The geometries were further validated by comparing to the very accurate MBPol force field~\cite{Babin2013/10.1021/ct400863t}; a detailed analysis can be found in the SI.

Population analyses of the gas-phase configurations were performed with the B3LYP density functional and the augmented cc-pVTZ Dunning basis~\cite{Dunning1992/10.1063/1.462569} (water dimer and fluoride-water complex) or the def2-QZVPPD basis~\cite{Peterson2003/10.1063/1.1622924,Rappoport2010/10.1063/1.3484283} (iodide-water complex) using Gaussian 09~\cite{g09} where the basis sets were chosen to give basis set converged results (see SI).
For the AIMD trajectories, configurations every \SI{5}{\femto\second} were analyzed, whereas for the AI-PIMD simulations at \SI{25}{\kelvin} configurations every \SI{25}{\femto\second} of every 10$^{\text{th}}$ bead were used. Configurations of the AI-PIMD simulations at \SI{300}{\kelvin} were selected from every 5$^{\text{th}}$ bead every \SI{5}{\femto\second}.
Note that we applied a standard hydrogen bond criterion (cutoff of 3.5~\AA{} on the oxygen-oxygen distance and 130\degree{} on the O-H-O angle) for the water dimer at \SI{300}{\kelvin} to discard non-bonded configurations caused by thermal fluctuations.

Bader's quantum theory of atoms in molecules (QTAIM) charges~\cite{Bader1991/10.1021/cr00005a013} were calculated with the AIMAll program~\cite{AIMAll} from the wavefunction provided by Gaussian 09, Hirshfeld charges were calculated directly with Gaussian 09 and natural population analysis (NPA) charges with calls to the NBO 6.0 program~\cite{Glendening2013/10.1002/jcc.23266}.
SAPT(2+3) interaction energies, including the extension for charge transfer~\cite{Hohenstein2010/10.1063/1.3451077,Stone2009/10.1016/j.cplett.2009.03.073}, were obtained using Psi4~\cite{Turney2012/10.1002/wcms.93} with the aug-cc-pVTZ basis set and the corresponding auxiliary basis set for density fitting.

To analyze electronic properties in liquid water including an explicit treatment of nuclear quantum effects, AI-PIMD and AIMD trajectories obtained in our previous study~\cite{Wang2014/10.1063/1.4894287} were used. The electronic structure was evaluated with the Quickstep module of the CP2K package~\cite{VandeVondele2005/10.1016/j.cpc.2004.12.014} using the BLYP-D3 functional~\cite{Becke1988/10.1103/PhysRevA.38.3098,Lee1988/10.1103/PhysRevB.37.785,Grimme2010/10.1063/1.3382344} and a DZVP basis. The i-PI program~\cite{Ceriotti2014/10.1016/j.cpc.2013.10.027} was used and in the AI-PIMD simulations each atom was represented by 6 beads using the PIGLET algorithm~\cite{Ceriotti2012/10.1103/PhysRevLett.109.100604}.

For the analysis of the fluoride ion solvated in water, AI-PIMD and AIMD simulations of the ion with 64 water molecules in periodic boundary conditions with a box size of 12.414~\AA{} were run for \SI{27}{\pico\second} and \SI{40}{\pico\second}, respectively, with a plane wave cutoff of 280~Ry and settings otherwise identical to those for the liquid water simulations.

To obtain the molecular charges in the condensed phase (both neat water and hydrated fluoride), the full electronic structure of the box was calculated for snapshots taken from our trajectories.
Configurations of the classical AIMD trajectory and of one bead of the AI-PIMD trajectory were analyzed every \SI{12.5}{\femto\second} using the Quickstep module of the CP2K package~\cite{VandeVondele2005/10.1016/j.cpc.2004.12.014}.
Stricter electronic structure settings were used, based on thorough testing of the convergence of the charges (see SI).
Specifically, the BLYP-D3 functional was kept and the SCF convergence was increased to $\epsilon_{\rm{SCF}}$=\num{1.e-10}, while the cutoff for the plane-wave representation of the charge density was set to 900~Ry.
The m-TZV2P basis set optimized for molecular systems (molopt)~\cite{VandeVondele2007/10.1063/1.2770708} was employed and the three population analyses also used for the gas-phase systems were performed in the following way.
To obtain Bader charges, the electronic density on a grid was analyzed with the Bader Charge Analysis program of the Henkelman group~\cite{Tang2009/10.1088/0953-8984/21/8/084204}.
Hirshfeld charges were determined directly with CP2K.
NPA charges are not available directly from CP2K, but were obtained by writing out the atomic orbital overlap and density matrix of every configuration and preparing NBO 6.0~\cite{Glendening2013/10.1002/jcc.23266} input.

To enable the decomposition of charge populations on molecules in liquid water, individual charge transfer contributions were calculated for hydrogen bonded dimers extracted from the bulk geometries.
For this, hydrogen bonded configurations were detected by applying a cutoff on the oxygen-oxygen distance $r_\text{OO}$ of 3.5~\AA{}, which corresponds to the first minimum of the radial distribution function, and a cutoff on the hydrogen bond angle $\measuredangle_{\text{OHO}}$ of 130\degree. 
Overall, \num{20000} dimer configurations were randomly extracted from the quantum ensemble, since it features a wider range of structures than the classical ensemble.

Population analyses for extracted dimer configurations were performed using the BLYP-D3 functional with otherwise similar specifications as for the gas-phase dimers.
We repeated the population analyses also using MP2 to ensure that the CT is not an artifact of DFT, obtaining results comparable to those from DFT (see Fig.~S12 in the SI for details). 

\begin{figure*}[th]
    \centering
    \includegraphics[width=1\textwidth]{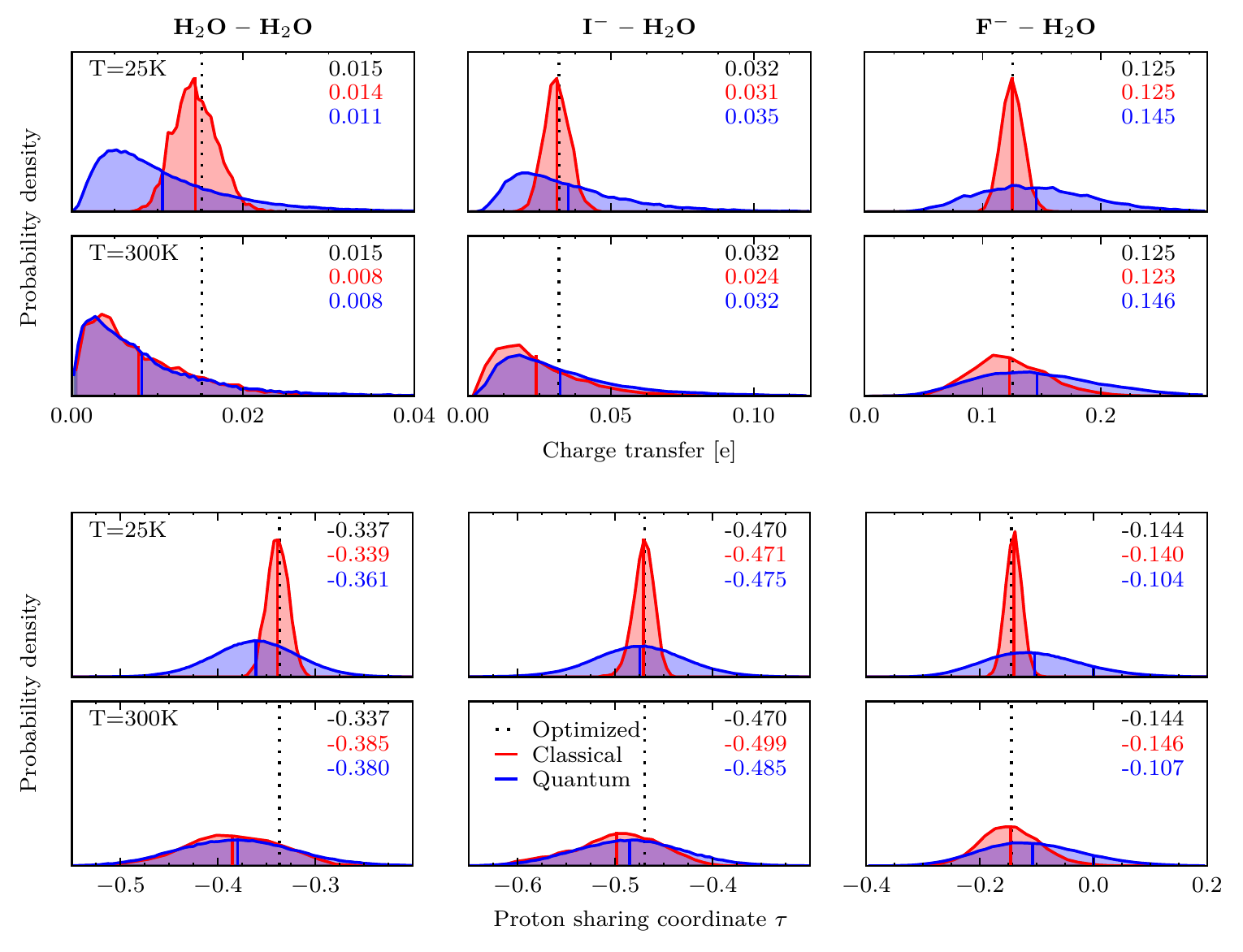}
    \caption{\label{fig:dimer_ct}
    Distributions of NPA charge transfer (top 6 panels) and distributions of the proton sharing coordinate $\tau$ (bottom 6 panels) for the water dimer (left), iodide-water (middle) and fluoride-water complex (right) from AIMD (Classical, red) and AI-PIMD (Quantum, blue) simulations at \SI{25}{\kelvin} and \SI{300}{\kelvin} as well as for the minimum energy structure (Optimized, dotted).
    The average of each distribution is marked with a solid line and printed in the respective color.
    The x- and y-axis ranges are identical for each pair of panels corresponding to the two temperatures.}
\end{figure*}

\section{Results}

In what follows we investigate CT in several hydrogen-bonded systems over a range of temperatures in both the gas and condensed phase. The water dimer was chosen as it is the prototypical hydrogen bond, the iodide-water complex was chosen as an intermediate strength hydrogen bond that is dominated by the charge of the diffuse and polarizable iodide and the fluoride-water complex was studied as an example with a very strong hydrogen bond.
First we consider these systems in the gas phase at the low temperature of 25~K as well as at ambient temperature to isolate the ground state contribution to CT. Second, to understand molecular charge populations in the condensed phase, we investigate CT in liquid water and around the hydrated fluoride anion. Finally, we show that it is possible to construct a simple model of CT based on proton sharing in hydrogen bonds.

For clarity, in the main text we report only NPA charge populations, but results obtained with QTAIM and the Hirshfeld method are reported in the SI. It should be noted that although the absolute amount of charge varies with the different population analyses, the trends and conclusions we demonstrate are universal.

\subsection{Gas-phase systems}

We start by examining the amount of charge transfer in the three gas phase complexes --- water-water, water-iodide, and water-fluoride --- at temperatures of \SI{25}{\kelvin} and \SI{300}{\kelvin}. At \SI{25}{\kelvin}, the properties of all the systems are dominated by the vibrational ground state. Hence, the influence of NQEs when minimal thermal fluctuations are present can be assessed. 
The results of the NPA CT analysis are shown in the top 6 panels of Fig.~\ref{fig:dimer_ct}, which compares the classical and quantum ensembles as well as the minimum energy structure. 
The CT is largest for the strongly hydrogen-bonded fluoride-water complex, less substantial for the iodide-water complex, which is of intermediate strength, and smallest for the neutral water dimer. 
At \SI{25}{\kelvin}, the AIMD simulations, which include only classical thermal fluctuations, show a narrow symmetric distribution of CT around an average which is very close to the value obtained for the minimum energy structure.
In contrast, when NQEs are included in the description of the system using AI-PIMD simulations, the distributions become substantially broader and more asymmetric with a long tail of high CT values. This shifts the average CT away from that obtained for the minimum energy structure. 
However, the direction of this shift differs for the three complexes, with the water dimer average CT decreased by 27~\%, while the fluoride-water complex CT is increased by 16~\%. 
These results thus show that, due to the asymmetry and the shift of the average in the quantum simulations, the minimum energy structure does not provide an adequate approximation to the quantum mechanical ground state. 

Upon increasing the temperature to 300~K, the difference between the classical and quantum CT distributions markedly decreases. This is largely due to changes in the classical distributions, while the quantum distributions are essentially unchanged from those at 25~K, suggesting that even at 300~K, CT is dominated by the ground state contribution. 
While for the water dimer the classical and quantum distributions look essentially identical, in the case of the two halide-water complexes, there are small residual differences, with the quantum average matching the minimum structure in the case of iodide and the classical average matching it for fluoride.
It is important to note, however, that given the shape and spread of the distributions, these matches should be considered entirely coincidental and should not be relied on in general.

\begin{figure}[t]
	\centering
	\includegraphics[width=0.9\linewidth]{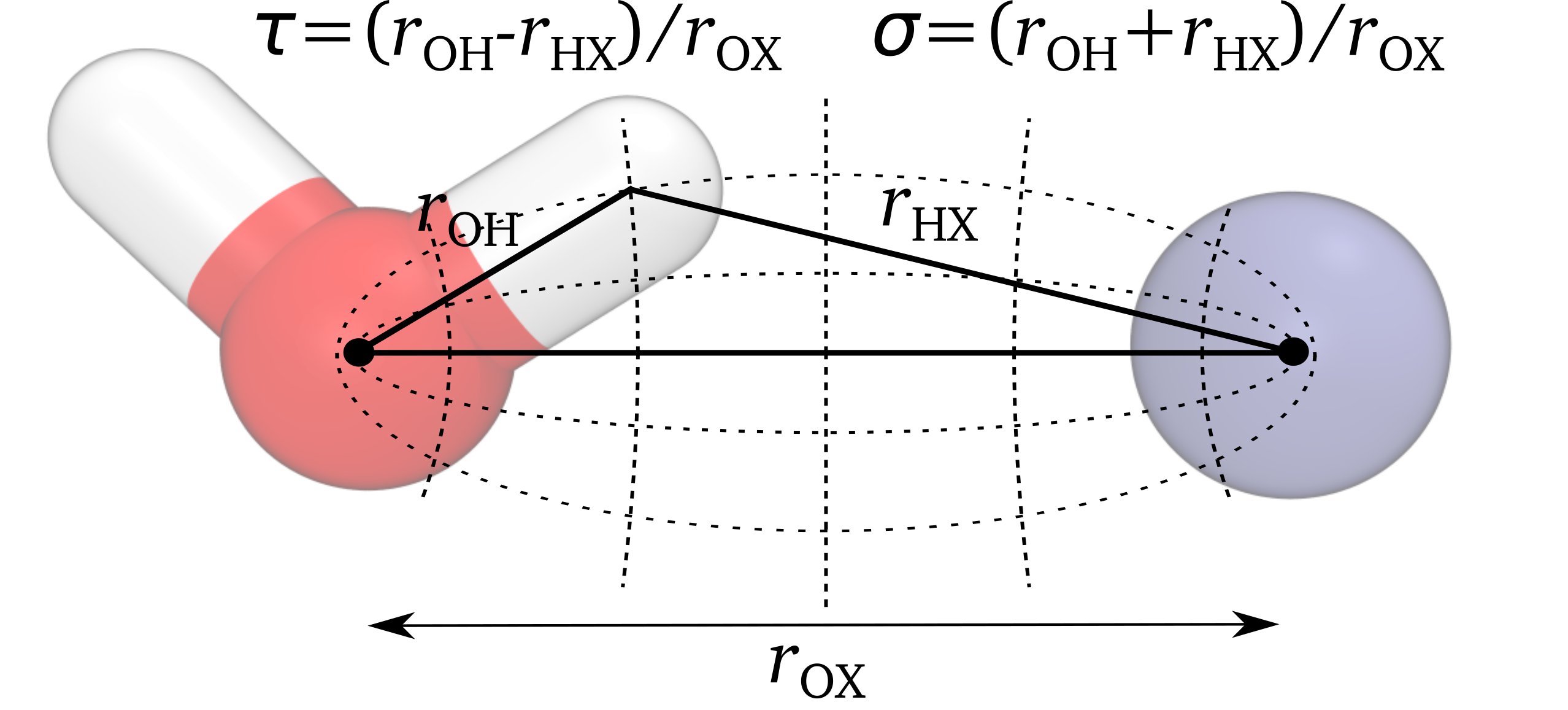}
    \caption{\label{fig:coord}
    Elliptical coordinate system for the hydrogen bond reduced to the two heavy atoms (O and X) and the hydrogen bonded hydrogen atom (H). The two orthogonal coordinates $\tau$ and $\sigma$ fully describe the hydrogen bond in this system together with the distance between the two foci of the ellipses which corresponds to the heavy-heavy atom distance $r_\text{OX}$. The close relation between the usual proton coordinate $\delta=r_\text{OH}-r_\text{HX}$ and $\tau$ should be noted.}
\end{figure}

To assess the structural changes which modulate the CT, we consider the elliptical coordinates defined in Fig.~\ref{fig:coord}.
These describe the position of the proton in the hydrogen bond in terms of a proton sharing coordinate $\tau$, a dimensionless version of another commonly used proton sharing coordinate~\cite{Voth1996/10.1063/1.470962,Marx1999/10.1038/17579,Ceriotti2013a/10.1073/pnas.1308560110}, and an orthogonal coordinate $\sigma$ which describes the deviation from linearity.
A higher value of $\tau$ corresponds to a hydrogen bond with the proton closer to the acceptor heavy atom, with $\tau$=0 representing a proton being shared equally between the two heavy atoms. As shown in Fig. S6, this coordinate strongly correlates with the amount of charge transfer.
This arises since, when the proton is more shared, the electron density shifts more towards the acceptor molecule, resulting in an increase in charge transfer.
As depicted in Fig.~\ref{fig:dimer_ct}, the changes of the distribution of $\tau$ (bottom 6 panels) are thus mirrored in the behaviour of the CT distribution (top 6 panels).
At the lower temperature, the distribution of $\tau$ is spread out considerably upon including NQEs for all systems.
In addition, for the water-dimer, NQEs decrease the average value of $\tau$, while for iodide-water it remains unchanged and for fluoride-water the average of $\tau$ is increased.
At ambient temperature, the classical and quantum distributions become almost identical for these gas phase systems, again matching the behavior of the magnitude of CT.

\begin{figure}[t]
    \centering
    \includegraphics[width=1\linewidth]{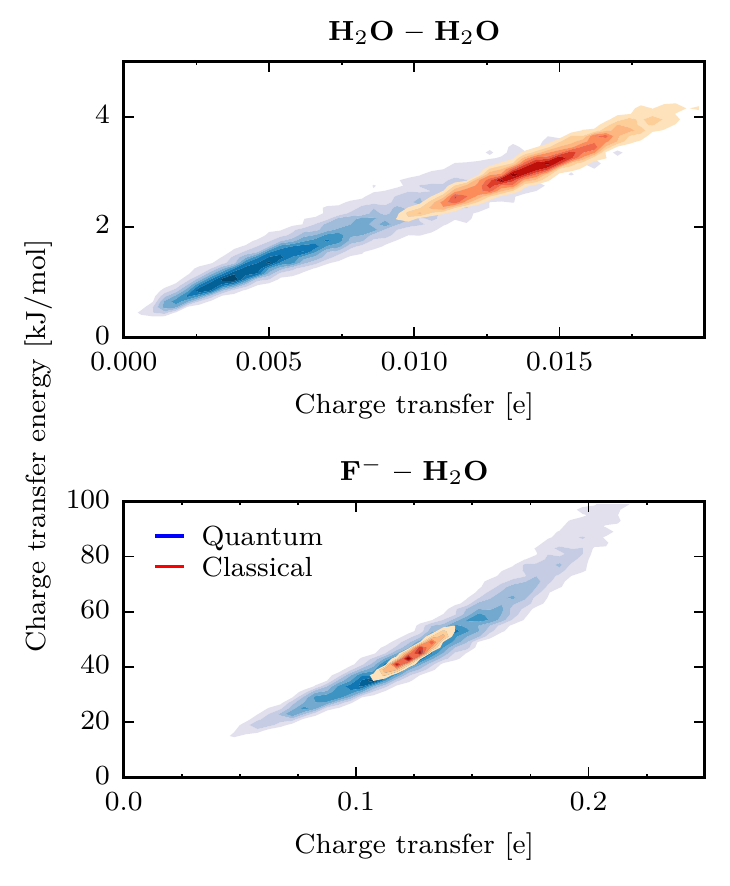}
    \caption{\label{fig:cor_npa_sapt}
    Correlation of NPA charge transfer with SAPT(2+3) charge transfer energy for the water dimer (top) and fluoride-water complex (bottom) from AIMD (Classical, red) and AI-PIMD (Quantum, blue) simulations at \SI{25}{\kelvin}.}
\end{figure}

Thus far, our analysis was based on the partitioning of the electronic density to determine the amount of electrons partially transferred through the hydrogen bond. 
To provide a measure of how this CT affects the binding energy of the complex, we performed calculations for the water dimer and fluoride-water complexes using the symmetry adapted perturbation theory (SAPT) method with an extension to partition the induction energy into charge transfer and polarization contributions.
Fig.~\ref{fig:cor_npa_sapt} shows that the charge transfer energy is strongly correlated with the amount of CT.
Overall, the CT energy is in the range of $\sim$0--4~kJ~mol$^{-1}$ for the water dimer, which corresponds to 15~\% of the total binding energy.
As already implied by the much larger amount of CT, the range of CT energies for the fluoride-water complex is over one order of magnitude larger, $\sim$20--100~kJ~mol$^{-1}$, and makes up roughly 30~\% of the total binding energy. Owing to this correlation, the CT energies are influenced by NQEs in a way analogous to the amount of CT and the proton sharing coordinate.

\subsection{Condensed phase}

\begin{figure}[t]
  \centering
  \includegraphics[width=1\linewidth]{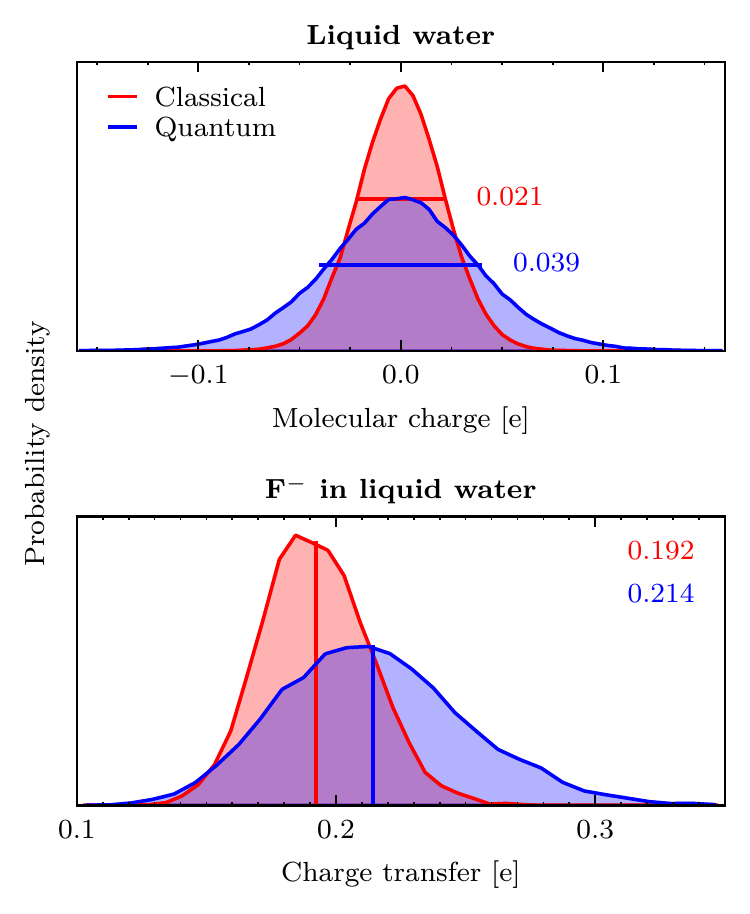}
  \caption{\label{fig:liquid_ct}
  Distribution of the molecular charge in liquid water (top) and charge transfer of the fluoride-ion solvated in water (bottom) obtained with the NPA method from AIMD (Classical, red) and AI-PIMD (Quantum, blue) simulations at \SI{300}{\kelvin}. The standard deviation (top) / average (bottom) of each distribution is marked with a solid line and stated in the respective color.}
\end{figure}

We now consider molecular CT in the condensed phase for ambient temperature liquid water as well as the hydrated fluoride anion.
The distributions of molecular NPA charge populations for these two systems with a classical and quantum description of the nuclei are depicted in Fig.~\ref{fig:liquid_ct}, while results for the other population analyses are shown in Fig.~S9.
In the case of the fluoride anion, CT only occurs in one direction --- electronic density is transferred from the fluoride anion to the surrounding water molecules through the hydrogen bonds that they donate to the ion.
Any deviation of charge population on the fluoride anion from the formal charge of -1 is thus the amount of CT due to all hydrogen bonds it receives in the liquid.
The hydrated fluoride anion in the liquid phase has an average CT of 0.19 e in the classical case.
This is only $\sim$50~\% higher than for the monohydrated anion in the gas phase, despite the increase in the number of hydrogen bonds it receives from 1 to an average of $\sim$4.5--5.0.
NQEs increase the average CT by 11~\% (slightly less than the 18~\% observed in the gas phase) and markedly increase the range of values observed towards higher charge transfer. 
Water molecules, unlike the fluoride anion, can both accept and donate hydrogen bonds. Due to the symmetry of the hydrogen bonding in the bulk liquid, the distribution of molecular charges is symmetric and has an average of zero and thus only the fluctuations can vary.
Our simulations show that NQEs increase these fluctuations around zero charge --- the quantum distribution in the top panel of Fig.~\ref{fig:liquid_ct} is considerably wider than the classical distribution.

\subsection{Analysis of charge transfer in liquid water}

Based on our observations above, we now demonstrate how one can provide a simple but accurate model of CT through the hydrogen bond. Previous investigations of charge transfer between water molecules in the liquid phase have modelled the molecular charge distribution by using a procedure in which formation of a hydrogen bond contributes a constant amount of charge transfer~\cite{Lee2011/10.1063/1.3589419,Vacha2011/10.1021/ja202081x,Vacha2011/10.1021/jz2014852,Rick2016/10.1002/jcc.24426}. In these studies, each molecule was assigned a hydrogen bond donor-acceptor balance $\Delta_\text{DA}$, calculated as the difference between the total number of donor $N_\text{D}$ and the total number of acceptor hydrogen bonds $N_\text{A}$ of a given molecule, $\Delta_\text{DA} = N_\text{D} - N_\text{A}$. This direct charge transfer model then gives a linear relationship between $\Delta_\text{DA}$ and the charge population on the molecule~\cite{Lee2011/10.1063/1.3589419}. However, as shown in Ref.~\citenum{Vacha2011/10.1021/jz2014852} and in Fig.~S10 for our liquid water trajectories, while the presence of a hydrogen bond shifts the charge population on a water molecule, that alone is not sufficient to explain the CT, as seen from the width and overlap of the distributions in Fig.~S10. Indeed, using this criterion there would be essentially no change expected in the CT upon including NQEs because the number of hydrogen bonds is only minimally perturbed. This suggests that the CT is determined by other parameters not captured by the binary criterion of hydrogen bond existence. Motivated by the strong dependence of charge transfer on the proton sharing coordinate $\tau$ in our gas-phase results (Fig.~\ref{fig:dimer_ct} and Fig.~S6) we show how incorporating this coordinate in a condensed-phase model of CT yields much improved predictions.

\begin{figure}[t]
    \centering
    \includegraphics[width=1\linewidth]{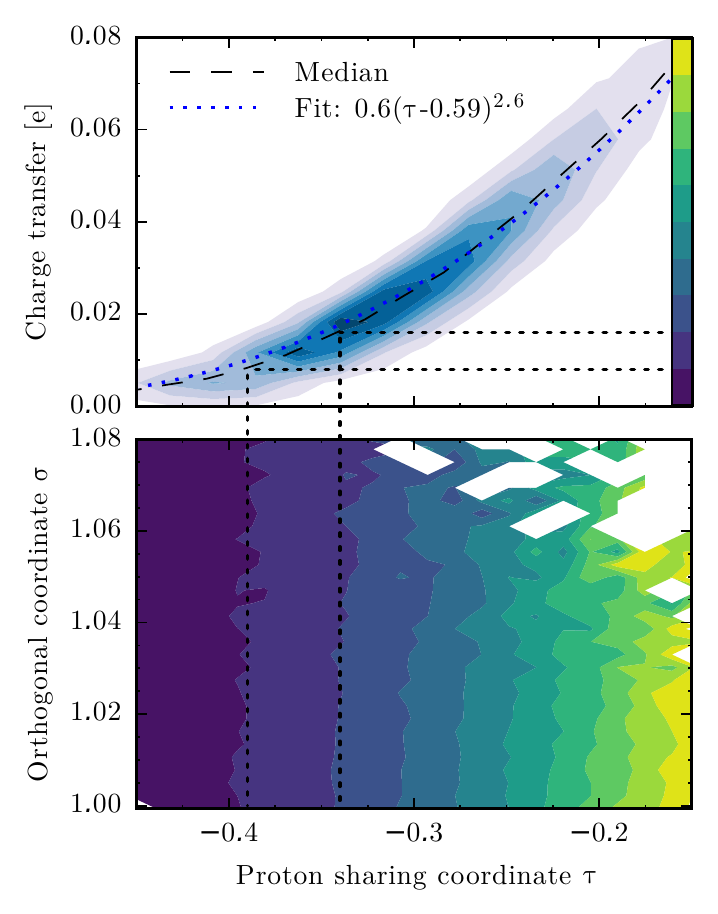}
    \caption{\label{fig:fit}
    The upper panel shows the correlation of the proton sharing coordinate $\tau$ with the charge transfer obtained with the NPA method for extracted dimers of the AI-PIMD simulation at \SI{300}{\kelvin}, where the probability is given in blue in equidistant steps and the median of the distribution is plotted as a dashed line, while a fit to the data is shown dotted in blue. The lower panel shows the charge transfer as a function of $\tau$ and the orthogonal coordinate $\sigma$ again in equidistant steps over the charge transfer range given by the upper panel. The dotted black lines spanning both panels should illustrate how the color map of the lower panel is to be interpreted.}
\end{figure}

To parametrize such a model for CT we begin by extracting \num{20000} dimers from our quantum simulations, which were chosen as they sample a wider range of CT values (see top panel of Fig.~\ref{fig:liquid_ct}). We then calculated the CT for each of these dimers separately to quantify the CT between that pair of water molecules. The resulting CT values obtained are plotted against $\tau$ in the top panel of Fig.~\ref{fig:fit}.
As for the dimer configurations obtained from the gas phase simulations, these show a strong correlation of the CT with $\tau$.
On the other hand, the CT shows no dependence on the coordinate $\sigma$ orthogonal to $\tau$, as shown in the bottom panel of Fig.~\ref{fig:fit}, suggesting that including this coordinate in the CT model would offer little benefit.
This relationship can be used to determine the expected amount of CT given a value of $\tau$.
To fit the dependence we used the following simple functional form that uses only three parameters
\begin{align}\label{eq:CT}
    \Delta q(\tau) = \left\{
    \begin{array}{ll}
        a\cdot\left(\tau-b\right)^{c} & \quad \text{if}\quad \tau > b \\
        0 & \quad \text{if}\quad \tau \le b
    \end{array}
    \right.,
\end{align}
where the values of the parameters $a$, $b$ and $c$ listed in Tab.~\ref{tab:para} were determined by a least squares fit to the data. When this function is used to predict the CT of the extracted dimer set, a correlation of the predicted charge with the charge from the population analysis of 96.4~\% is observed. This high correlation coefficient shows that the other degrees of freedom account only for under 4~\% of the CT and that a simple functional form is thus able to describe the dependence appropriately.

\begin{figure}[t]
    \centering
    \includegraphics[width=1\linewidth]{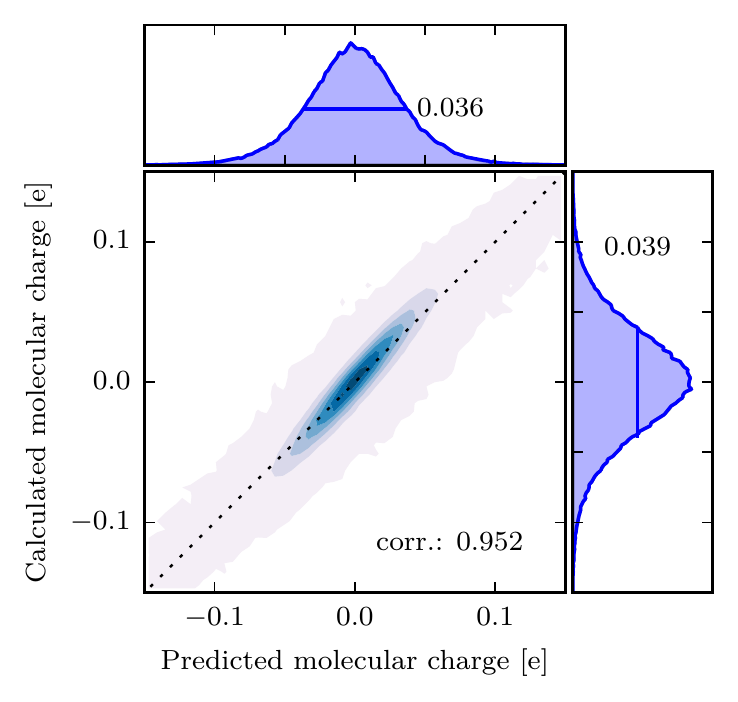}
    \caption{\label{fig:cor_pred_calc}
    Correlation of the predicted molecular charge using the obtained relation between charge transfer and the instantaneous value of the proton sharing coordinate $\tau$ with the calculated molecular charge for the NPA method of the PI-AIMD simulations of liquid water at \SI{300}{\kelvin} as well as the corresponding distributions of the molecular charge with stated standard deviation.}
\end{figure}

With this model, the total charge on each molecule in the bulk liquid can be predicted by identifying the hydrogen bond donors and acceptors of each molecule and their $\tau$ values. One can then calculate the charge $q_i$ of water molecule $i$ by summing the incoming and outgoing CT contributions,
\begin{align}
  q_i = \sum_{n = 1}^{N_\text{D}} \Delta q(\tau_n) - \sum_{m = 1}^{N_\text{A}} \Delta q(\tau_m),
\end{align}
where $\Delta q(\tau)$ is obtained from Eq.~\ref{eq:CT}.
Fig.~\ref{fig:cor_pred_calc} shows the predictions from this simple model compared to the molecular charges obtained from a full electronic structure calculation for the quantum liquid water ensemble.
Excellent agreement is obtained for the NPA data shown, with the model giving slightly smaller molecular charges than the full electronic structure calculations.
For NPA, the prediction works the best with a correlation coefficient of 95.2~\%, while QTAIM and the Hirshfeld method yield correlation coefficients of 86.2~\% and 83.6~\%.
The prediction for the classical ensemble shows comparably high correlation coefficients of 93~\%, 80~\% and 81~\%, respectively.
This shows that the prediction of molecular charges based solely on the proton sharing coordinate $\tau$ in hydrogen bonds is a good approximation for the total charge population in the liquid phase.

\begin{table}[t]
    \centering
    \caption{\label{tab:para}
    Parameters of the non-linear least square fit of charge transfer as a function of the proton sharing coordinate $\tau$ for NPA, Bader's QTAIM and the Hirshfeld method assuming the functional form given by Eq.~\ref{eq:CT}.}
    \vspace{10pt}
    \begin{tabular}{@{} l *3d @{}}
    \toprule[1.0pt]
     & \multicolumn{1}{r}{NPA}  & \multicolumn{1}{r}{QTAIM}  & \multicolumn{1}{r}{Hirshfeld} \\
    \toprule[1.0pt]
    $a$ & 0.6254 &  0.2107 &  0.4872 \\
    $b$ &-0.5932 & -0.6464 & -0.6526 \\
    $c$ & 2.5937 &  1.8677 &  1.4376 \\
    \bottomrule[0.5pt]
    \end{tabular}
\end{table}

\begin{figure}[t]
    \centering
    \includegraphics[width=1\linewidth]{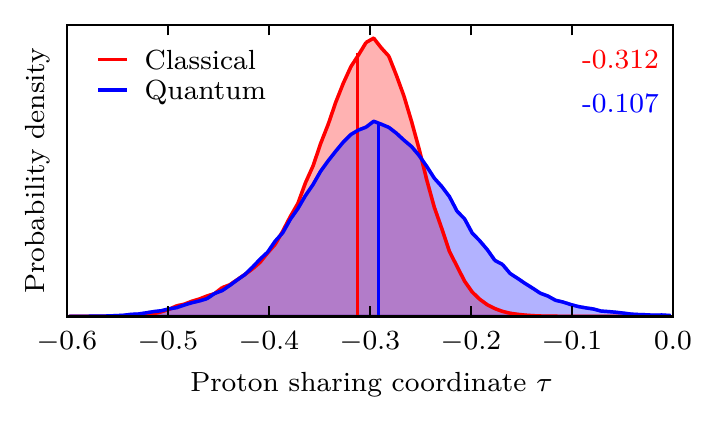}
    \caption{\label{fig:tau_bulk}
    Distributions of the proton sharing coordinate $\tau$ in liquid water from classical and quantum simulations.}
\end{figure}

Since we have demonstrated that CT and the resulting molecular charge in liquid water are almost entirely dictated by a single geometric parameter of the hydrogen bond, we can now elucidate the difference between the quantum and classical ensembles. Fig.~\ref{fig:tau_bulk} shows the distributions of this geometric parameter, the proton sharing coordinate $\tau$ for liquid water. Although the classical and quantum ensembles match closely at more negative values of the coordinate, as one approaches zero, corresponding to a highly shared proton, the quantum distribution exhibits significantly higher probability density. This increase in proton sharing, observed in a number of previous studies ~\cite{Ceriotti2013a/10.1073/pnas.1308560110,Wang2014/10.1063/1.4894287, Ceriotti2016/10.1021/acs.chemrev.5b00674,Marsalek2016/10.1063/1.4941093,Marsalek2017/1702.07797}, is vital in this case, as it solely determines the amount of CT. This is in stark contrast to many other properties which exhibit a cancellation between quantum effects along $\tau$ and the orthogonal coordinate $\sigma$ \cite{Manolopoulos2009/10.1063/1.3167790,McKenzie2014/10.1063/1.4873352,Michaelides2011/10.1073/pnas.1016653108}.

\section{Conclusions}

In conclusion, by comparing quantum and classical simulations over a wide range of hydrogen bonded structures in both the gas and condensed phases we have shown that NQEs can make large qualitative and quantitative changes to the magnitude of CT. We have demonstrated that these results are robust with respect to the choice of electronic structure method and population analysis approach.
In particular, the insensitivity of CT to the coordinates orthogonal to the hydrogen bond direction means that the higher CT due to increased proton sharing is not compensated by decreases due to other degrees of freedom. Due to the large amount of zero point energy in the $\tau$ coordinate in the quantum case, the additional thermal energy at the higher temperature only changes the position distribution by a small amount, in contrast to the classical case.
This leads to a very small temperature dependence of the quantum CT distribution compared to that observed classically. 
As such, one can easily be misled by considering classical configurations or minimum energy structures alone when assessing the amount of CT.

Further, we were able to show that it is possible to accurately model the charge population on a water molecule in the condensed phase by considering only the sharing of protons with its immediate hydrogen bonding neighbors. This approach is able to explain 80--95~\% of the charge population on a given molecule even though it lacks higher body effects, showing that the wider environment plays only a minor role. Indeed, recent work has shown that 2-body effects are sufficient to model the dipole moment surface of liquid water~\cite{Liu2015/10.1063/1.4921045}, further suggesting that the electrostatics of water are well captured at this level. Our analysis thus provides physical insight into the origin of charge transfer in hydrogen bonded systems as well as a simple basis for its accurate modelling.

\begin{acknowledgments}
This material is based upon work supported by the National Science Foundation under Grant No. CHE-1652960. T.E.M also acknowledges support from a Cottrell Scholarship from the Research Corporation for Science Advancement. We would like to thank Stanford University and the Stanford Research Computing Center for providing computational resources and support that have contributed to these research results. C.S. gratefully acknowledges financial support from a scholarship from the German Academic Scholarship Foundation (\textit{Studienstiftung des Deutschen Volkes}).
\end{acknowledgments}

\pagebreak

\end{document}